\documentclass[conference]{IEEEtran}
\IEEEoverridecommandlockouts
\usepackage{cite}
\usepackage{amsmath,amssymb,amsfonts}
\usepackage{algorithmic}
\usepackage{booktabs}
\usepackage[acronym]{glossaries}
\usepackage{graphicx}
\usepackage{hyperref}
\usepackage{orcidlink} 
\usepackage{url}
\usepackage{tabularx}
\usepackage{textcomp}
\usepackage{xcolor}
\usepackage{threeparttable} 

\usepackage{tikz}
\usetikzlibrary{arrows.meta, shapes, positioning}

\usepackage{booktabs}

\usepackage{paper_acronyms}

\def\BibTeX{{\rm B\kern-.05em{\sc i\kern-.025em b}\kern-.08em
    T\kern-.1667em\lower.7ex\hbox{E}\kern-.125emX}}

\begin{document}
\bstctlcite{BSTcontrol} 

\title{
    Experimental Analysis of FreeRTOS Dependability through Targeted Fault Injection Campaigns
    \thanks{
        This work was supported by project COLTRANE-V funded by the Ministero dell'Università e della Ricerca within the PRIN 2022 program (D.D.104 - 02/02/2022) and carried out within the Space It Up project funded by the Italian Space Agency, ASI, and the Ministry of University and Research, MUR, under contract n. 2024-5-E.0 - CUP n. I53D24000060005.
    }
}

\author{
    \IEEEauthorblockN{
        Luca Mannella~\orcidlink{0000-0001-5738-9094}, 
        Stefano Di Carlo~\orcidlink{0000-0002-7512-5356},
        Alessandro Savino~\orcidlink{0000-0003-0529-7950}, 
    }
    \IEEEauthorblockA{
        \textit{Department of Control and Computer Engineering, Politecnico di Torino},
        Turin, Italy \\
        \{name.surname@polito.it\}
    }
}

\maketitle

\begin{abstract}

\glspl{rtos} play a crucial role in safety-critical domains, where deterministic and predictable task execution is essential. Yet they are increasingly exposed to ionizing radiation, which can compromise system dependability.
To assess FreeRTOS under such conditions, we introduce KRONOS, a software-based, non-intrusive post-propagation \gls{fi} framework that injects transient and permanent faults into \gls{os}-visible kernel data structures without specialized hardware or debug interfaces. Using KRONOS, we conduct an extensive \gls{fi} campaign on core FreeRTOS kernel components, including scheduler-related variables and \glspl{tcb}, characterizing the impact of kernel-level corruptions on functional correctness, timing behavior, and availability. The results show that corruption of pointer and key scheduler-related variables frequently leads to crashes, whereas many \gls{tcb} fields have only a limited impact on system availability.

\end{abstract}


\glsresetall 

\section{Introduction}
\label{sec: introduction}

Real-time embedded systems have become the backbone of modern critical infrastructures, extending far beyond consumer electronics to underpin applications in automotive control, aerospace and satellite systems, and other platforms~\cite{Chen:2015uu}. Their pervasive deployment in safety and mission-critical domains demands stringent guarantees of dependability, encompassing reliability, safety, and fault tolerance~\cite{Solouki2024, Aalund2025}.
To meet the diverse and demanding requirements of these applications, embedded devices typically rely on specialized software stacks. In some cases, this software is implemented as tightly coupled, application-specific firmware, while in others, it is built atop \glspl{rtos} that orchestrate complex scheduling, resource management, and timing constraints.

Despite advances in design, embedded systems remain vulnerable to both internal faults, such as hardware defects, software bugs, and malicious tampering, and external threats from their operational environment~\cite{Solouki2024}. Notably, transient and permanent faults, especially in space and high-altitude applications, have a significant impact, and their occurrence is increasing even at ground level as device geometries shrink~\cite{Srour2013Review, Vallero:2019aa, Portolan:2019, Papadimitriou:2021tf, Aalund2025}. 
In this context, radiation-induced effects such as \glspl{seu} and persistent hardware faults may propagate to kernel-visible state and compromise correct execution~\cite{Srour2013Review}.
The growing vulnerability of modern embedded platforms to these phenomena has stimulated the development of robust reliability assessment and mitigation techniques~\cite{Aalund2025}.

Among the available methodologies, \gls{fi} is considered a cornerstone for evaluating system resilience. By deliberately introducing faults (either in hardware, software, or at the system level), researchers can systematically study fault propagation, error detection, and recovery mechanisms under controlled conditions~\cite{Benso2011, De-Sio:2023aa}. While many \gls{fi} studies focus on the effects of faults at the application level, some works also emphasize the importance of considering the \gls{rtos} kernel and its data structures, e.g., scheduler queues, timers, and \glspl{tcb}, since faults in this layer can significantly influence system behavior and safety~\cite{De-Sio:2023aa, Solouki2024}. Nonetheless, a systematic analysis of \glspl{rtos} under transient and permanent faults, and of their implications for system dependability, remains an open research direction.

This paper investigates how transient and permanent faults affecting key \gls{rtos} kernel data structures influence system dependability once these faults propagate to the \gls{rtos}-visible state. The main contributions are:
(1) design of \gls{KRONOS}, a novel, software-based, \gls{fi} framework for FreeRTOS~\cite{Guan2016FreeRTOS};
(2) extensive kernel-level \gls{fi} campaign on FreeRTOS, for both transient and permanent fault models in RTOS-visible kernel state.
(3) detailed analysis of outcome distributions and identification of critical kernel data structures, facilitating early-stage vulnerability assessment and identifying the most sensitive FreeRTOS components. 

The remainder of this paper is organized as follows: Section~\ref{sec: background} reviews relevant related work on \gls{fi}. Section~\ref{sec: proposed-approach} details the proposed \gls{fi} methodology and its integration with FreeRTOS. Section~\ref{sec: setup} describes the experimental setup, while Section~\ref{sec: results} presents and discusses the results of our \gls{fi} campaigns. Finally, Section~\ref{sec: conclusion} concludes the paper.

\section{Related Work}
\label{sec: background}
\glsunset{kronos}

\gls{fi} is a widely used technique for assessing system dependability by purposefully introducing faults into the system and observing its behavior under these engineered fault scenarios~\cite{Kanawati:1995aa, Carreira:1998aa, Ebrahimi:2014tj, doi:10.1049/PBCS057E, Bodmann:2021vs}.
\gls{fi} can be classified (by injection mechanism) into four categories: (i) physical, (ii) hardware-based, (iii) software-based, and (iv) model-based.
\emph{Physical} \gls{fi} involves subjecting the system implementation to external phenomena, such as radiation.
\emph{Hardware-based} techniques exploit hardware interfaces, e.g., the debug interface, to modify register contents or memory.
\emph{Software-based} methods alter the software layer, targeting variables or memory buffers.
Finally, \emph{model-based} \gls{fi} operates on abstract models, such as \glspl{hdl} or instruction-set simulators, enabling \gls{fi} campaigns in simulation environments. More details on those techniques can be found in~\cite{doi:10.1049/PBCS057E, Bodmann:2021vs}.

Targeting the \gls{os} layer for \gls{fi} is particularly challenging~\cite{Bosio:2022aa, Casseau:2021aa}, since faults affecting core kernel data can lead to unrecoverable errors and system hang. While similar effects may also occur when injecting faults at the application level, they tend to appear less frequently, and managing hangs is typically more time-consuming, adding complexity to \gls{fi} campaigns~\cite{De-Sio:2023aa}.

Recent advances in model-based \gls{fi} frameworks include gem5-MARVEL~\cite{Chatzopoulos2024gem5MARVEL}.
Built on the gem5 microarchitectural simulator~\cite{Binkert2025gem5}, it enables \glspl{fi} at a detailed hardware level across multiple \glspl{isa} and system components, including \gls{os} kernels within simulation environments.
While it offers fine-grained microarchitectural fault modeling and \gls{os}-level \gls{fi} capability, the approach entails long simulation times and complexity. More broadly, microarchitectural \gls{fi} comes at the expense of long simulation times.

Prior work has extensively investigated \gls{fi} at the hardware and microarchitectural levels, often highlighting the limitations of simplified high-level simulation approaches. Papadimitriou and Gizopoulos~\cite{Papadimitriou:2021tf} performed \gls{fi} within a processor's microarchitecture and demonstrated that purely software-level \glspl{fi} techniques that flip bits in program variables can overlook essential microarchitectural effects, sometimes even reversing vulnerability rankings when compared to full-stack analyses. Similarly, Cho et al.\cite{10.1145/2463209.2488859} examined architecture- and memory-level \gls{fi} methods, showing that while such techniques offer high execution speed, their error-rate estimates may deviate by an order of magnitude. Collectively, these studies underscore the inherent limitations of high-abstraction \glspl{fi} in capturing realistic hardware error behavior and fail to consider fault-handling mechanisms at the \gls{os} level.
		
In contrast, our work targets the \gls{os} layer by injecting faults directly into \gls{rtos} kernel memory structures to evaluate the dependability of system software itself. By deliberately abstracting from microarchitectural modeling, the proposed approach focuses on analyzing the \gls{os} response to corrupted internal states—for instance, the behavior of task schedulers, kernel queues, and memory managers under fault conditions. Rather than estimating hardware fault rates, our objective is to uncover \gls{os}-level failure modes and robustness limitations that remain unobservable through hardware-centric \gls{fi}. This perspective complements existing cross-layer fault analyses by elucidating how faults propagate and manifest within the \gls{os}.

However, other studies began targeting \gls{os} structures. 
\gls{fiat}~\cite{Barton:1990aa} was one of the earliest software-based \gls{fi} tools targeting the kernel. It applied abstract data-corruption models to running tasks but lacked the resolution to selectively corrupt internal \gls{os} structures, such as the \glspl{tcb}.
Another microkernel-based framework is MAFALDA~\cite{Arlat:2002DoCm}. Even if it can inject faults into API arguments or microkernel memory segments, its architecture, which separates injection and execution across machines, offers limited insight into internal fault propagation.
\gls{rtos} Guardian~\cite{Silva:2011aa} introduced a hardware-based fault monitoring architecture for embedded \gls{rtos} systems. Its components non-intrusively monitor the system bus to validate \gls{rtos} scheduling integrity.
Although effective in electromagnetic-induced faults detection per IEC 61000-4-29~\cite{IEC61000EMC}, its analysis is limited to temporal anomalies and does not extend to corruption of memory-resident kernel structures.

Mamone et al.~\cite{Mamone:2020ta} presented one of the first experimental analyses targeting the reliability of FreeRTOS kernel structures through \gls{fi}. That work demonstrated the relevance of \glspl{fi} directly into \gls{rtos} 's internal data structures to identify critical components, using a hardware-based experimental setup and focusing on transient-fault effects. Our work aims at extending this analysis along several dimensions, as detailed in Sections~\ref{sec: proposed-approach}, \ref{sec: setup}, and \ref{sec: results}. In particular, \gls{kronos} enables automated, repeatable kernel-level \gls{fi} without dedicated hardware, and supports both transient and permanent fault models.
Moreover, the proposed approach extends the experimental scope to a broader set of kernel objects and execution conditions, enabling a more comprehensive characterization of \gls{rtos}-level failure modes.

\section{Proposed Approach}
\label{sec: proposed-approach}

\gls{kronos} is a post-propagation \gls{fi} framework that operates on \gls{os}-visible memory, assuming faults have already reached kernel-accessible state. This abstraction allows controlled analysis of \gls{rtos} behavior under corrupted kernel data without modeling lower-level propagation. It systematically reveals the most vulnerable kernel structures and failure modes, providing a conditional vulnerability assessment of FreeRTOS components rather than absolute hardware fault rates.

\begin{figure}[htb]
	\centering
    \includegraphics[width=0.70\linewidth]{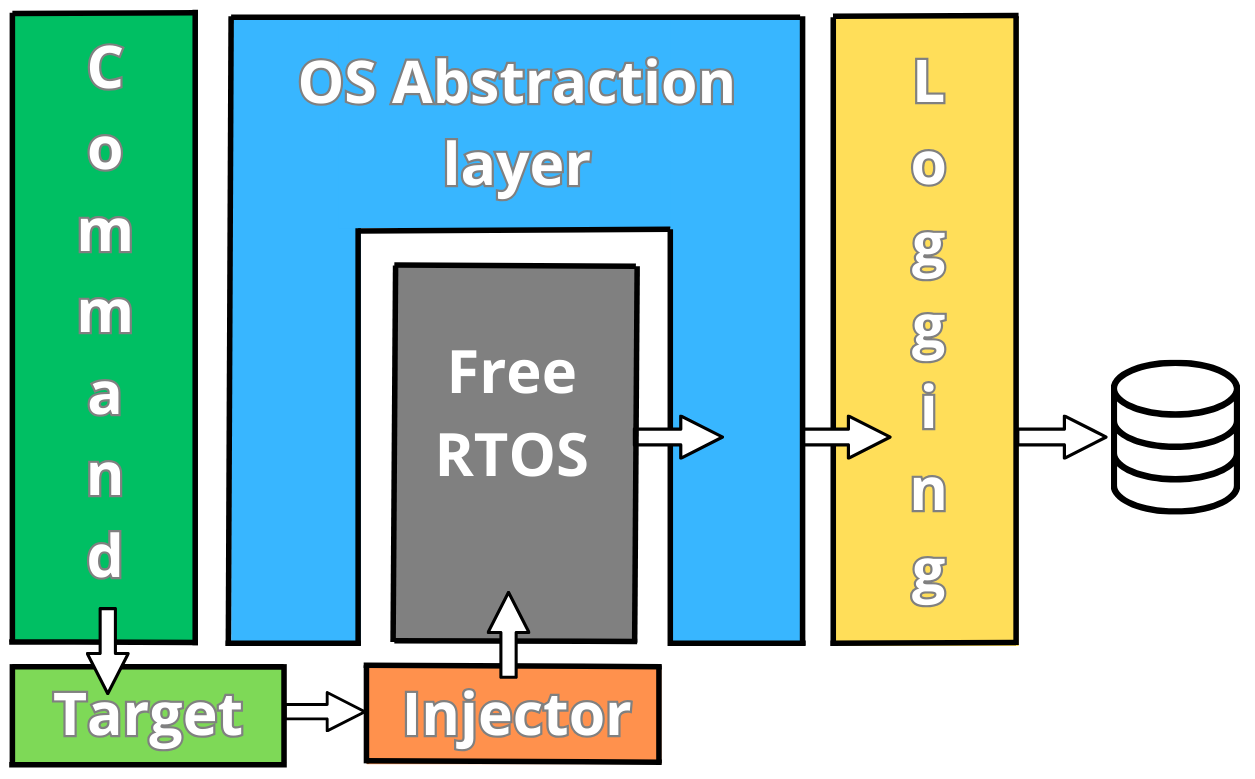}
    \caption{Overview of the main components of \gls{kronos} and their interactions.}
	\label{fig: kronos_arch}
\end{figure}

\gls{kronos} is built on the FreeRTOS port for Linux and Windows, allowing us to run a complete FreeRTOS system as a set of user-space processes and threads on a hosted operating system without hardware.
Leveraging the capability of FreeRTOS to run as a system process, \gls{kronos} is fully software-based and requires no architectural modifications or external debug interfaces. 
Its portability and direct access to kernel structures make it well-suited for systematic dependability evaluation of FreeRTOS-based systems.
Specifically, \gls{KRONOS} is composed of: (i) a \textit{command} module, which specifies injection commands; (ii) a \textit{target} module, responsible for extracting and acquiring the injection targets; (iii) an \textit{\gls{os}-abstraction layer}, allowing the injector to execute on different FreeRTOS ports; (iv) an injector module; and (v) a logging module, which hooks into FreeRTOS at key events (e.g., task switches) to record runtime behavior. Figure~\ref{fig: kronos_arch} shows a high-level overview of the main components of the \gls{fi} application.

\subsection{Fault Injection and Result Collection Workflow}
 
In its simulation ports, FreeRTOS runs as a process (and the port implementation guarantees the tick reference). Hence, \gls{kronos} is designed to run in a dedicated thread within the same process. This design allows the injector thread to share the same process address space as the simulated FreeRTOS, without altering the FreeRTOS implementation and configuration. This shared-memory model enables precise access to \gls{fi} within key FreeRTOS components, with minimal impact on the system and without requiring special hardware or architectural modifications.

The \gls{fi} workflow comprises four key steps:

(\textbf{I}) \emph{Injection Target Specification}: Qualified kernel objects (variables, pointers, lists, structures) are identified at runtime via specialized gatherer functions through name, base memory address, size, and structural hierarchy within kernel data.

(\textbf{II}) \emph{Golden Run Profiling}: FreeRTOS tasks execute uninterrupted to establish timing baselines and expected outputs for functional correctness comparison.

(\textbf{III}) \emph{Run Initialization}: The injector loads target parameters (injection time offset from scheduler start, byte/bit offsets, fault type) alongside the golden run profile, starts the FreeRTOS scheduler (allowing normal task/timer execution), and sleeps until the specified injection time.

(\textbf{IV}) \emph{\acrlong{fi}}: The injector wakes and applies the configured fault.

Post-injection, \gls{kronos} monitors the scheduler state: if the scheduler is functional (only the \emph{Idle task} is ready), it lets the Idle task hook gracefully terminate the system; otherwise, it forces a shutdown after a configurable timeout. Logs and outputs are then analyzed against golden-run references for task correctness and execution timing, in line with broader software test monitoring approaches already proposed~\cite{10.1145/2463209.2488859}.

Each run can be classified with one of the following standard outcome categories:

\begin{itemize}
    \item \textbf{Benign}: The system completes its task within the expected time and produces correct results.  
    \item \textbf{DELAY}: The result is functionally correct, but the execution exceeds the predefined deadline (which led to a performance or real‐time violation).
    \item \textbf{\gls{sdc}}: The system finishes on time but yields an incorrect result without raising errors. To better capture the nuances of real-time execution, this class has two labels: if \gls{sdc} is produced with a delay, it is labeled accordingly (\gls{sdc} DELAY).
    \item \textbf{HANG}: The system never completes (no output or completion signal), typically due to deadlock, livelock, or an unhandled fault. The time limit is configurable as a percentage of the golden run time.
    \item \textbf{CRASH}: The system terminates abruptly (e.g., via exception, abort, or unhandled signal), preventing normal shutdown and output generation.
    \item \textbf{INVALID}: Since certain \gls{os} structures are valid targets only under specific use-case behaviors, we classify them into a separate category when, even though injections are attempted on the target, it is not considered valid.
\end{itemize}

\subsection{Fault Injector Implementation}

\gls{kronos} implements two software-level fault models on \gls{rtos}-visible kernel state: transient faults and permanent faults. These models do not emulate the physical generation of radiation-induced effects at the circuit level; rather, they represent the condition in which a fault has already propagated to a kernel data structure and become architecturally visible to the \gls{rtos}. In this sense, the proposed \gls{fi} approach is post-propagation and targets the software manifestation of the fault.

\textbf{Transient faults} model a non-persistent single-event corruption of a kernel target, represented as a one-time bit flip in the runtime memory of the selected kernel object. At the configured \gls{fi} instant, the injector directly accesses the kernel process memory space, locates the target memory address plus the byte and bit offsets, and performs the flip of the targeted bit. After the injection, subsequent kernel writes to the same location are not altered. This model captures a temporary fault effect that becomes visible in the \gls{rtos} state but is not permanently retained.

\textbf{Permanent faults} model a persistent corruption of a kernel target, represented as a stuck-at condition on a selected bit of the targeted kernel object. Since the corrupted value must be preserved across subsequent updates, \gls{kronos} introduces a compile-time patching mechanism that preprocesses the FreeRTOS kernel code before building the hosted executable. The patcher uses PCRE2 to locate all writes to the targeted locations and inserts a bitwise mask enforcing the stuck-at condition after each write. This strategy ensures that the corrupted bit remains permanently forced during execution, at the cost of a minimal extra execution time. It therefore models a persistent software-visible fault condition rather than the low-level physical mechanisms that originally generated it.

\section{Experimental Setup}
\label{sec: setup}

To evaluate FreeRTOS's dependability, we conducted an extensive \gls{fi} campaign across different kernel structures. Specifically, the targets of our analysis were divided into four groups: 
(i) \emph{Global Variables:} variables that influence scheduling decisions and help manage FreeRTOS, described in Table~\ref{tab: FreeRTOSGlobalVarsTargets};
(ii) \emph{Pointers:} pointers to FreeRTOS global variables or data structures used to manage the \gls{rtos} (see Table~\ref{tab: FreeRTOSPointerTargets});
(iii) \emph{Lists:} structures used by FreeRTOS for managing tasks. Tasks can belong to different lists based on their state, i.e., ready, delayed, pending, or suspended (see Table~\ref{tab: FreeRTOSListTargets});
(iv) \emph{Current TCB:} is the data structure associated with the currently running task (see Table~\ref{tab: FreeRTOSCurrentTCBTargets1}). It contains relevant information about the task, i.e., task state, priority, and stack pointer.

\begin{table}[tb]
    \centering
    \caption{FreeRTOS Global Variable Targets}
    \label{tab: FreeRTOSGlobalVarsTargets}
    \resizebox{\columnwidth}{!}{%
    \begin{tabular}{l l p{0.50\linewidth}}
        \toprule
        \textsc{Target} & \textsc{Type} & \textsc{Description} \\
        \midrule
        \texttt{uxCurrentNumberOfTasks} & \textit{UBaseType\_t} & Number of active tasks. \\
        \texttt{uxDeletedTasksWaitingCleanup} & \textit{UBaseType\_t} & Number of TCB pending cleanup by the IDLE task. \\
        \texttt{xPendedTicks} & \textit{TickType\_t} & Ticks accumulated while scheduler was suspended. \\
        \texttt{uxTaskNumber} & \textit{UBaseType\_t} & Counter for task creation (unique ID). \\ 
        \texttt{uxTopReadyPriority} & \textit{UBaseType\_t} & Highest priority level with at least one ready task. \\
        \texttt{xNextTaskUnblockTime} & \textit{TickType\_t} & Tick count when the next delayed task is due to unblock. \\ 
        \texttt{xTickCount} & \textit{TickType\_t} & System tick counter since scheduler start. \\
        \texttt{xNumOfOverflows} & \textit{BaseType\_t} & Number of times \texttt{xTickCount} has wrapped around. \\ 
        \texttt{xSchedulerRunning} & \textit{BaseType\_t} & Scheduler started flag. \\
        \texttt{xTimerQueue} & \textit{QueueHandle\_t} & Handle of the internal timer command queue. \\
        \texttt{xTimerTaskHandle} & \textit{TaskHandle\_t} & Timer Daemon task handle. \\ 
        \texttt{xYieldPending} & \textit{BaseType\_t} & Context switch pending flag \\ 
        \bottomrule
    \end{tabular}
    } %
\end{table}

\begin{table}
    \centering
    \caption{FreeRTOS Pointer Targets.}
    \label{tab: FreeRTOSPointerTargets}
    \begin{tabularx}{\linewidth}{lX}
        \toprule
        \textsc{Target} & \textsc{Description} \\
        \midrule
        \texttt{pxCurrentTCB} & Pointer to the TCB of the running task. \\
        \texttt{pxCurrentTimerList} & Pointer to the current timer list. \\
        \texttt{pxDelayedTaskList} & Pointer to the delayed task list. \\
        \texttt{pxOverflowDelayedTaskList} & Pointer to the overflowed delayed task list. \\
        \texttt{pxOverflowTimerList} & Pointer to the current overflowed timer list. \\
        \texttt{xIdleTaskHandle} & Pointer to the Idle task's TCB \\
        \bottomrule
    \end{tabularx}
\end{table}

\begin{table}
    \centering
    \caption{FreeRTOS List Targets.}
    \label{tab: FreeRTOSListTargets}
    \begin{tabularx}{\linewidth}{lX}
        \toprule
        \textsc{Target} & \textsc{Description} \\
        \midrule
        \texttt{pxReadyTasksLists} & Array of ready-task lists indexed by task priority. \\ 
        \texttt{xDelayedTaskList1} & Lists of tasks delayed via vTaskDelay(). \\ 
        \texttt{xDelayedTaskList2} & Lists of tasks delayed via vTaskDelay(). \\ 
        \texttt{xPendingReadyList} & Tasks that became ready while the scheduler was locked. \\ 
        \texttt{xActiveTimerList1} & List of active software timers. \\ 
        \texttt{xActiveTimerList2} & List of active software timers. \\
        \texttt{xSuspendedTaskList} & List of suspended tasks. \\ 
        \texttt{xTasksWaitingTermination} & List of deleted tasks pending resource cleanup by the Idle task. \\ 
        \bottomrule
    \end{tabularx}
\end{table}

\begin{table}
    \centering
    \caption{FreeRTOS current \gls{tcb} Targets. They can be accessed through \texttt{pxCurrentTCB}.}
    \label{tab: FreeRTOSCurrentTCBTargets1}
    \resizebox{\columnwidth}{!}{%
    \begin{tabular}{l l p{0.55\linewidth}}
        \toprule
        \textsc{Target} & \textsc{Type} & \textsc{Description} \\
        \midrule
        \texttt{pcTaskName} & \textit{char*} & Name of the current task. \\
        \texttt{pxStack} & \textit{StackType\_t*} & Pointer to the base of the current task stack. \\
        \texttt{pxTaskTag} & \textit{TaskHookFunction\_t} & Pointer to the application task tag of the currently running task. \\
        \texttt{pxTopOfStack} & \textit{StackType\_t*} & Pointer to the top of the stack of the currently executing task. \\
        \texttt{ucDelayAborted} & \textit{uint8\_t} & Flags whether a blocking delay was prematurely aborted for the current task. \\
        \texttt{ucNotifyState} & \textit{uint8\_t*} & Array of integer used by the task notification mechanism. \\
        \texttt{ulNotifiedValue} & \textit{uint32\_t*} & Array of integer used by the task notification mechanism. \\
        \texttt{ulRunTimeCounter} & \textit{uint32\_t} & Accumulates the task's time spent in the running state. \\
        \texttt{uxBasePriority} & \textit{UBaseType\_t} & Current task's base priority used by the mutex priority inheritance mechanism. \\
        \texttt{uxMutexesHeld} & \textit{UBaseType\_t} & Current task's number of mutexes held. \\
        \texttt{uxPriority} & \textit{UBaseType\_t} & Priority of the current task. \\
        \texttt{uxTaskNumber} & \textit{UBaseType\_t} & User-defined task tag. \\
        \texttt{uxTCBNumber} & \textit{UBaseType\_t} & Kernel-assigned unique TCB ID. \\ 
        \texttt{xEventListItem} & \textit{ListItem\_t} & List node for events (queues, semaphores). \\
        \texttt{xStateListItem} & \textit{ListItem\_t} & List node linking TCB into scheduler lists. \\
        \bottomrule
    \end{tabular}
    } %
\end{table}

The experimental setup employs five benchmarks from the TACLeBench suite executed by different tasks with different priorities~\cite{TACLeBench}: (i) \texttt{SHA} (classic hash function), (ii) \texttt{FFT} (Fast Fourier Transform computation), (iii) \texttt{CUBIC} (cubic equation solver), (iv) \texttt{HUFF\_DEC} (Huffman decoding), and (v) \texttt{ADPCM\_ENC} (adaptive pulse-code modulation encoding). These benchmarks cover cryptographic, signal-processing, and compression workloads, providing a diverse task mix. The first three tasks (\texttt{SHA}, \texttt{FFT}, and \texttt{CUBIC}) were configured with priority 1, \texttt{HUFF\_DEC} with priority 2, and \texttt{ADPCM\_ENC} with priority 3. Priorities were intentionally differentiated to exercise multiple \emph{ready lists} and create non-trivial preemption behavior during the campaign.
Each task executes its computation once and then self-deletes, ensuring the FreeRTOS scheduler can detect task completion and shut down properly.

The FreeRTOS kernel was configured with a 1 kHz tick rate, 7 priority levels, cooperative multitasking, dynamic task creation, and support for software timers, counting semaphores, recursive mutexes, and queue sets. Both transient and permanent faults were considered, with injections occurring within a 10,000 ns window after scheduler start (delay = 5\% deviation, hang = 300\% of golden run time). Each run executed the five TACLeBench tasks plus system/timer daemon tasks until graceful shutdown.

To achieve statistically meaningful results, we applied the binomial proportion estimation approach from~\cite{LeveugleR.2009SfiQ} (32-bit reference), treating the fault space as effectively infinite. This yielded 666 \gls{fi} injections per location (99\% confidence, 5\% margin of error), totaling 83,916 injections across all targets.

To assess execution-time stability, the \gls{fi} campaign was executed five times using four parallel worker processes. It was completed in an average wall-clock time of 76.7~s, with a standard deviation of 7.1~s (min--max: 70.3--91.3~s). These results confirm that \gls{kronos} can execute large-scale \gls{fi} campaigns within practical time bounds while maintaining repeatability across independent injection samples, making it ideal for early-stage development.

\section{Experimental Results}
\label{sec: results}

The experimental campaign demonstrates that, under \gls{os}-visible kernel memory corruptions, FreeRTOS exhibits moderate fault tolerance while several kernel components remain highly vulnerable (Table~\ref{tab: fi-campaign}). Approximately 70\% of runs completed successfully, about 3\% experienced delayed execution, and more than 20\% resulted in system crashes, indicating limited fault-handling capability in the default configuration. \gls{sdc} occurrences were infrequent (below 2\%) and generally associated with timing deviations, whereas fewer than 5\% of injections targeted invalid objects. 
Comparable failure rates were observed for transient and permanent faults across most kernel components, as corruptions of critical data structures frequently caused immediate or irrecoverable failures, such as scheduler inconsistencies or invalid memory accesses, so fault persistence often did not materially change the observed run-level outcome within a given execution.
These findings indicate that \gls{rtos} dependability is highly dependent on workload characteristics and operational context.

\begin{table}[ht]
    \centering
    \caption{Outcome of the whole \gls{fi} campaign divided by fault type.}
    \label{tab: fi-campaign}
    \resizebox{\columnwidth}{!}{%
    \begin{tabular}{llllllll}
        \toprule
        \textsc{Fault Type} & \textsc{Benign} & \textsc{Delay} & \textsc{SDC} & \textsc{SDC (Delay)} & \textsc{Hang} & \textsc{Crash} & \textsc{Invalid} \\
        \midrule
        Transient & 70.16\% & 2.80\% & 0.04\% & 1.69\% & 0.00\% & 20.43\% & 4.88\% \\
        Permanent & 69.66\% & 3.00\% & 0.00\% & 1.65\% & 0.00\% & 20.82\% & 4.88\% \\
        \bottomrule
    \end{tabular}
    } %
\end{table}

\begin{figure}
    \centering
    \includegraphics[width=0.90\columnwidth]{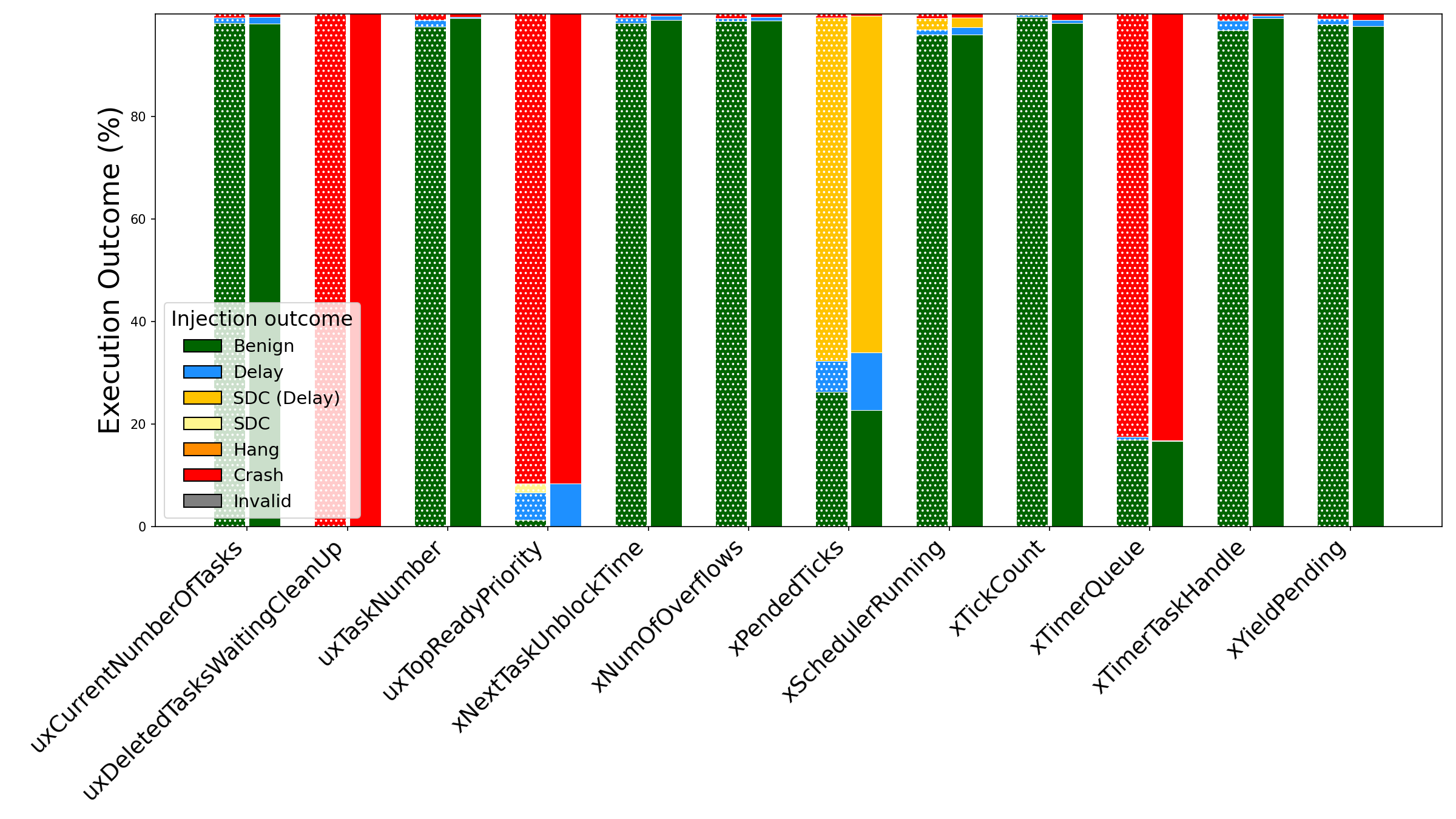}
    \caption{Injections on \emph{FreeRTOS} \textbf{variables}: transient (on the left) and permanent (on the right) faults.}
    \label{fig: results-variables}
\end{figure}

A category-based analysis shows that several \emph{variables}, such as \texttt{uxDeletedTasksWaitingCleanup}, \texttt{uxTopReadyPriority}, and \texttt{xTimerQueue}, exhibit near-total failure rates when altered, almost always resulting in system crashes. This highlights their critical role in core scheduler and resource-handling decisions.
Less sensitive variables, such as \texttt{xTickCount} or \texttt{uxCurrentNumberOfTasks}, typically result in correct execution but can occasionally cause crashes or delays, particularly under specific timing conditions. 
Notably, altering \texttt{xPendedTicks} produces a large percentage of \gls{sdc}, always associated with a delay (see Figure~\ref{fig: results-variables}).

\begin{figure}
    \centering
    \includegraphics[width=0.90\columnwidth]{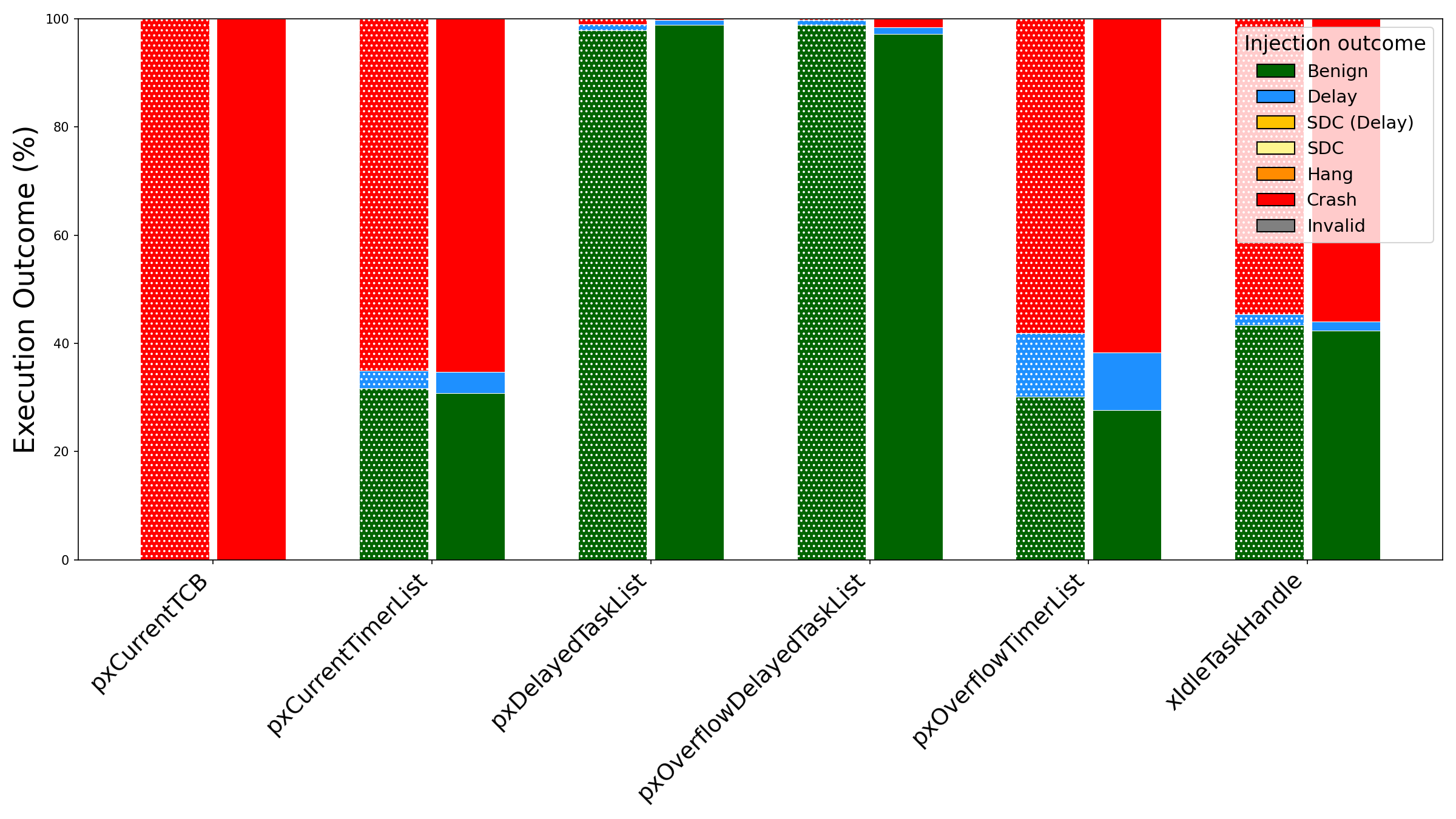}
    \caption{Injections on \emph{FreeRTOS} \textbf{pointers}: transient (on the left) and permanent (on the right) faults.}
    \label{fig: result-pointers}
\end{figure}

As expected, \emph{Pointer} variables are the most fault-sensitive elements: bit-flips almost always cause system crashes, as they can instantly invalidate objects or list references. Pointers to dynamic schedulers, delayed lists, or timer lists are the most sensitive, while \texttt{pxCurrentTCB}, whose corruption leads to crashes in all runs for both fault types, is the most critical.
Other sensitive elements are \texttt{pxCurrentTimerList} and \texttt{xIdleTaskHandle}, which, when altered, account for more than 50\% of system crashes.
Notably, affecting this variable category does not produce any \gls{sdc} (Figure~\ref{fig: result-pointers}).

\begin{figure}
    \centering
    \includegraphics[width=0.90\columnwidth]{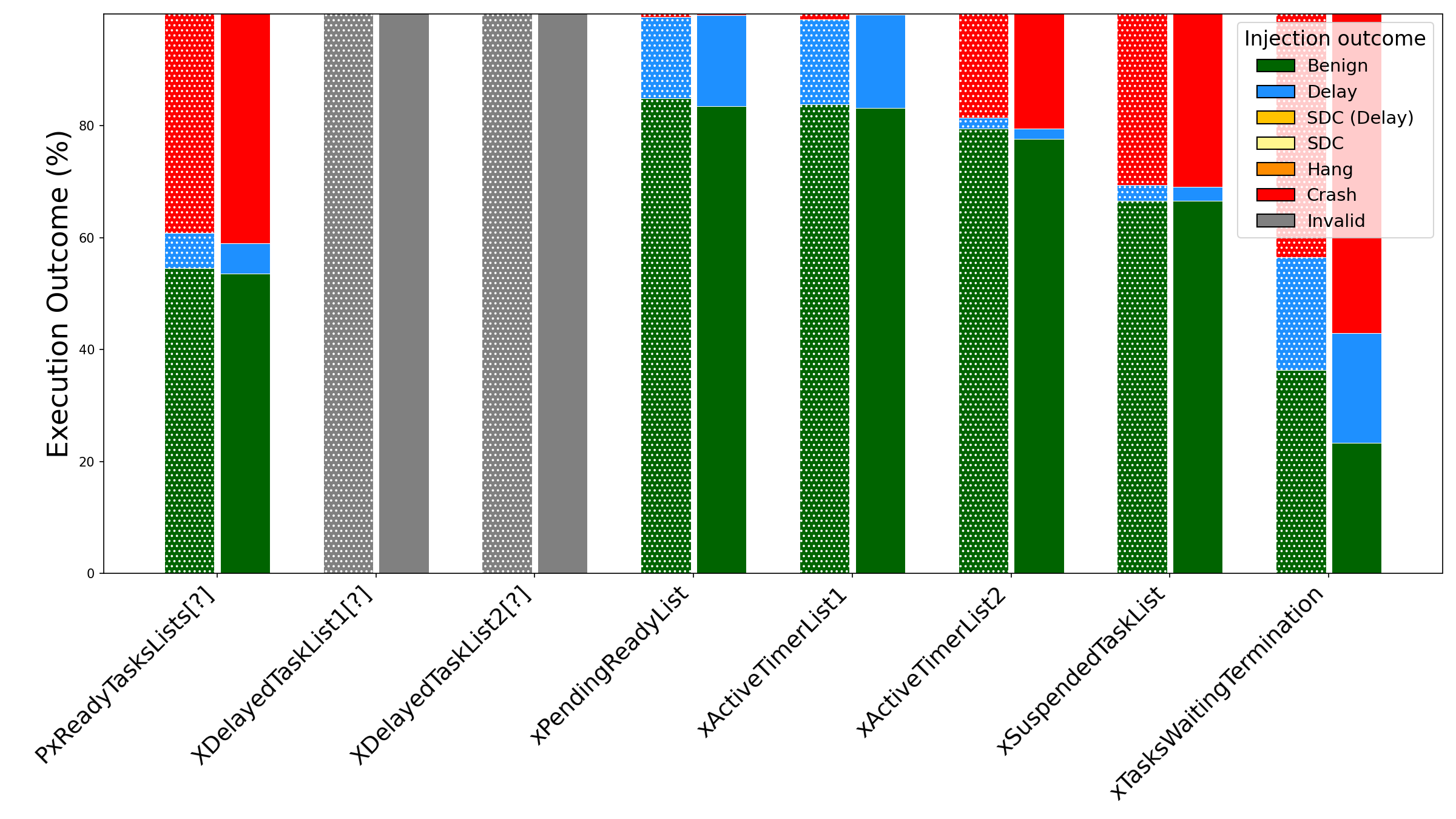}
    \caption{Injections on \emph{FreeRTOS} \textbf{lists}: transient (on the left) and permanent (on the right) faults.}
    \label{fig: results-lists}
\end{figure}

Faults injected in list objects lead to less frequent crashes compared with the other category (approximately from 20\% to 50\%), but with a small subset yielding delays (see Figure~\ref{fig: results-lists}). However, the crash rate in this category tends to increase for permanent faults. 
Injections into the task waiting termination lists (i.e., \texttt{xTasksWaitingTermination}) show exceptionally high sensitivity. In contrast, lists associated with less active kernel operations (e.g., \texttt{xSuspendedTaskList}) exhibit a higher rate of correct execution, even though the risk of a system crash remains non-negligible.
Notably, even affecting this variable category does not produce any \gls{sdc}.

As it is possible to observe in Figure~\ref{fig: results-lists}, some targets were never valid at injection time.
In particular, the delayed task lists were empty because the adopted use case is mostly computation-oriented and did not require tasks to enter delayed states at the selected injection instants.

\begin{figure}
    \centering
    \includegraphics[width=0.95\columnwidth]{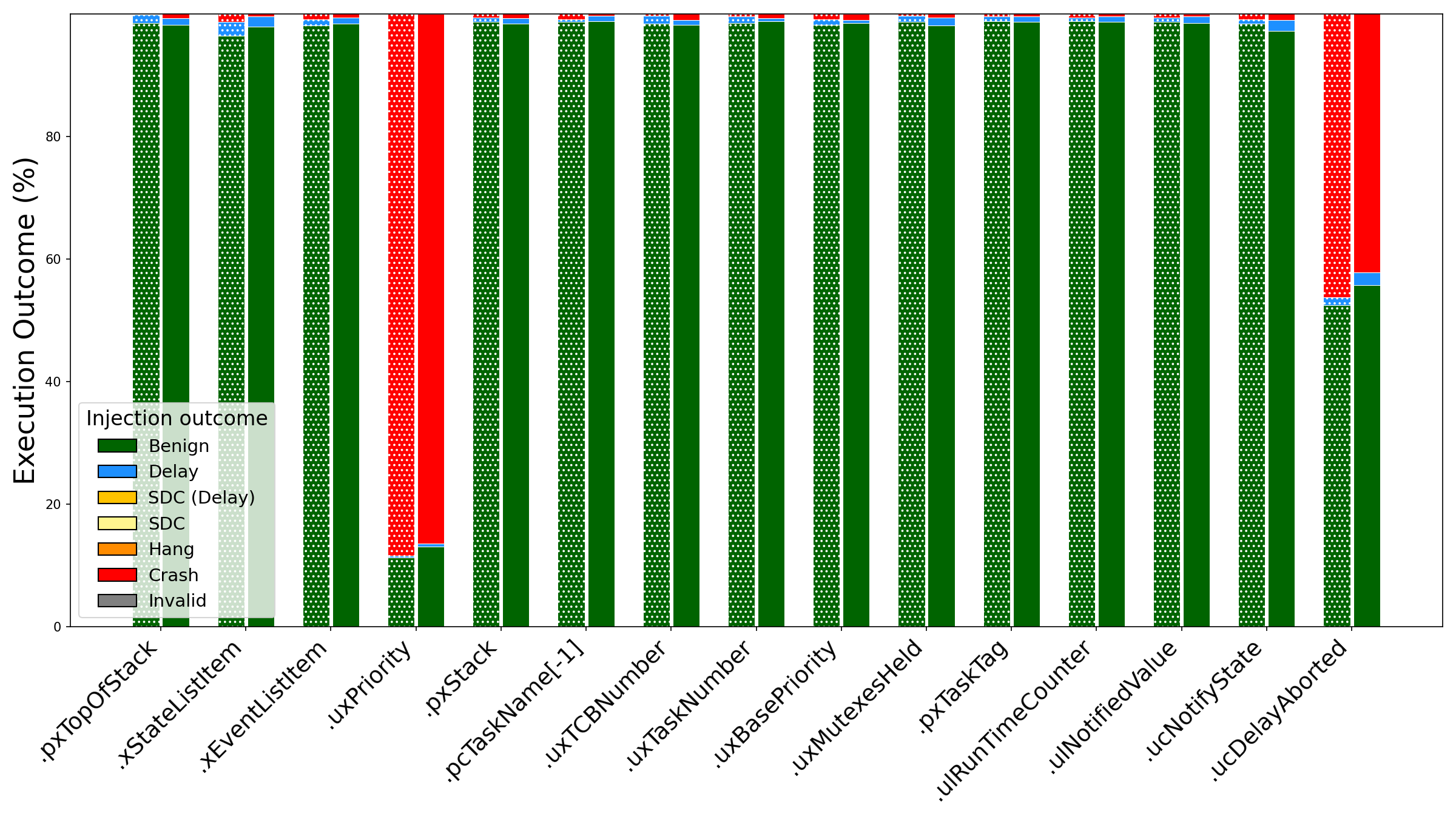}
    \caption{Injections on \emph{FreeRTOS} \textbf{current TCB} fields: transient (on the left) and permanent (on the right) faults.}
    \label{fig: results-TCB}
\end{figure}

Figure \ref{fig: results-TCB} reports that corrupting the fields inside the current \gls{tcb} did not produce particularly sensitive effects, with a small percentage of crashes and delays in most cases.
The most sensitive field is \texttt{uxPriority}, which, if altered, caused a crash in more than 80\% of runs. This evidence underscores the importance of priority management for deterministic \gls{rtos} operation.
Also, \texttt{ucDelayAborted} caused a crash in more than 40\% of runs when altered.
Notably, even affecting the fields of the TCB does not produce any \gls{sdc}.

\section{Conclusion}
\label{sec: conclusion}

This work extended earlier experimental analyses of FreeRTOS reliability using \gls{kronos}, a systematic software-based \gls{fi} framework targeting kernel-level data structures.
By operating on \gls{os}-visible memory state, \gls{kronos} enables repeatable evaluation of \gls{rtos} failure modes without specialized hardware or debug interfaces.

An extensive \gls{fi} campaign on critical FreeRTOS kernel objects showed that, although many corruptions are tolerated, core scheduler-related structures remain highly vulnerable, often causing crashes or deadline violations. The similar impact of transient and permanent faults indicates that corruption of critical \gls{rtos} state frequently leads to immediate, unrecoverable failures.
The analysis highlights the conditional vulnerability of FreeRTOS kernel components once faults become architecturally visible and complements lower-level fault-propagation studies, helping \gls{rtos} designers identify kernel structures needing stronger protection or monitoring in safety- and mission-critical systems.

Future work includes extending \gls{kronos} to other \glspl{rtos} and integrating cross-layer models that relate \gls{os}-visible faults to specific device-level phenomena.

\section*{Acknowledgment}
\small
The authors thank Giovanni De Florio and Dimitri Schiavone for their contributions to this research during their master's theses.

\label{sec:bib}
\bibliographystyle{IEEEtran}
\bibliography{bibliography}

\end{document}